\documentclass[structabstract]{aa}  
\usepackage{psfig,graphicx,epsfig}
\usepackage{lscape}
\usepackage{txfonts}
\usepackage{natbib}
\begin{document}

   \title{Study of HST counterparts to Chandra X-ray sources in the Globular Cluster M71}
   \titlerunning{Study of HST counterparts to Chandra X-ray sources in the Globular Cluster M71}
   \authorrunning{R. H. H. Huang et al.}
   \author{R. H. H. Huang\inst{1} \and W. Becker\inst{1} \and P. D. Edmonds\inst{2} \and 
              R. F. Elsner\inst{3} C. O. Heinke\inst{4}  \and B. C. Hsieh\inst{5}}
   \offprints{R. H. H. Huang (rhuang@mpe.mpg.de)}
   \institute{Max-Planck-Institut f\"ur extraterrestrische Physik,
               Giessenbachstrasse 1, 85748 Garching, Germany \and
               Harvard-Smithsonian Center for Astrophysics, Cambridge, MA 02138, USA \and 
               NASA Marshall Space Flight Center, Huntsville, AL 35812, USA \and
               Department of Physics, University of Alberta, Edmonton, Alberta, Canada \and
               Institute of Astronomy and Astrophysics, Academia Sinica,
               Taipei 10617, Taiwan }

  \abstract 
   {}
   {We report on archival Hubble Space Telescope (HST) observations of the globular cluster M71 (NGC 6838).}
   {These observations, covering the core of the globular cluster, were performed by the Advanced Camera for Surveys (ACS) 
   and the Wide Field Planetary Camera 2 (WFPC2). Inside the half-mass radius ($r_\mathrm{h}$ = 1$\farcm$65) of M71, 
   we find 33 candidate optical counterparts to 25 out of 29 Chandra X-ray sources while outside the half-mass radius, 
   6 possible optical counterparts to 4 X-ray sources are found.}
   {Based on the X-ray and optical properties of the identifications, we find 1 certain and 7 candidate cataclysmic 
   variables (CVs). We also classify 2 and 12 X-ray sources as certain and potential chromospherically active 
   binaries (ABs), respectively. The only star in the error circle of the known millisecond pulsar (MSP) is inconsistent 
   with being the optical counterpart.}
   {The number of X-ray faint sources with $L_\mathrm{X}>4\times10^{30}$~ergs~s$^{-1}$ (0.5--6.0\,keV) found 
    in M71 is higher than extrapolations from other clusters on the basis of either collision frequency or mass.
   Since the core density of M71 is relatively low, we suggest that those CVs and ABs are primordial in origin.}

   \maketitle
%________________________________________________________________

%%%%%% Section1 %%%%%

\section{Introduction}

There are 158 Galactic globular clusters (GGCs) found in the halo of our galaxy and typically 
contain $10^{4}$ -- $10^{7}$ stars. They are very old and dense star systems and tightly bound 
by gravity, which gives them their spherical shapes and relative high stellar density toward the center. 
The dense stellar environment in globular clusters triggers various dynamical interactions, ie. exchanges 
in encounters with binaries, direct collisions, destruction of binaries, and tidal capture. These dynamical 
interactions not only can change the evolution of individual stars, but also can produce tight binary 
systems \citep[see, e.g.,][for review]{1998ashman,2004verbunt}.

One of the most powerful ways to probe the binary content of globular clusters is by studying 
the X-ray source population. In the early 1970s, X-ray sources with luminosity greater than
$10^{36}~\mathrm{ergs~s^{-1}}$ were first detected by using the Uhuru and OSO-7 Observatories. Following the 
Einstein and ROSAT era, the number of faint X-ray sources ($L_\mathrm{X} < 10^{34.5}~\mathrm{ergs~s^{-1}}$)
was dramatically increased. Those bright X-ray sources have been identified with low-mass X-ray 
binaries \citep[LMXBs;][]{1984grindlay} while the identification of the weaker sources remained 
limited due to low photon statistics and insufficient spatial resolution. The launch of the Chandra X-ray 
Observatory ushered in a new age of studying the crowded centers of Galactic globular clusters
with a far greater sensitivity and resolving power than ever before \citep[e.g.,][]{2001grindlay, 2001grindlay2}. 
With the aid of the Hubble Space Telescope (HST), many of these faint X-ray sources were identified as 
quiescent low-mass X-ray binaries (qLMXBs; in which a neutron star accretes matter from its companion 
at a low rate), cataclysmic variables (CVs; in which a white dwarf accretes from its low-mass companion), 
and millisecond pulsars (MSPs), as well as chromospherically active binaries (ABs; e.g., RS CVn and BY Dra systems)
\citep[e.g., ][]{2001grindlay2, 2002pooley, 2003edmonds1, 2003edmonds2, 2005heinke, 2004bassa, 
2006kong, 2007lugger, 2008bassa}.

The globular cluster M71 (NGC\,6838) lies close to the Galactic plane with Galactic longitude $l=56\fdg74$ 
and latitude $b = - 4\fdg56$. Similarly to 47\,Tuc, it is a fairly metal-rich globular cluster with metallicity of 
[Fe/H] = $- 0.73$. Due to its relatively small distance (d $\sim$ 4\,kpc) to Earth and low central luminosity 
density ($\rho_\mathrm{c} = 10^{3.05}\,\mathrm{L_{\odot}/pc^{3}}$), M71 is a good target for both optical 
and X-ray observations. The core, half-mass, and tidal radii are $r_\mathrm{c}$ = 0$\farcm$63, 
$r_\mathrm{h}$ = 1$\farcm$65, and $r_\mathrm{t}$ =8$\farcm$96, respectively. M71 shows no evidence 
for core collapse. Its moderate optical reddening $E_\mathrm{B-V}$ = 0.25 may be converted into a nominal 
X-ray absorption column of $N_\mathrm{H} = 1.39 \times 10^{21} \mathrm{cm^{-2}}$ \citep{1995predehl}. 
The aforementioned parameters related to M71 were obtained from 
\citet[][ updated 2003\footnote{http://physwww.mcmaster.ca/~harris/Databases.html}]{1996harris}.

In this work we report on archival Chandra and HST observations of the globular cluster M71. 
We have obtained a 52.4-kilosecond Chandra observation of M71 taken with the Advanced CCD 
Imaging Spectrometer (ACIS), reaching the limiting X-ray luminosities of $1.5\times10^{30}~\mathrm{ergs~s^{-1}}$ 
and $7.9\times10^{29}~\mathrm{ergs~s^{-1}}$ in the energy ranges of 0.3--8.0 and 0.5--2.5\,keV, respectively. 
In \citet{2008elsner}, we reported the identification of 29 X-ray sources within the cluster half-mass radius, 
including the known millisecond pulsar PSR J1953+1846A (M71A), and their X-ray properties, and found 
that $18\pm6$ of these 29 sources are likely to be associated with M71 from a radial distribution analysis.
The present paper extends our study of the X-ray sources in M71 by using archival HST data to identify 
optical counterparts to the majority of M71's X-ray sources, improving our understanding of their nature.

In $\S$2 we discuss the Chandra X-ray observations and spectral analysis. HST observations, data reduction
and analysis are described in $\S$3. In $\S$4 we present the source identification. A discussion and 
comparison with other globular clusters is given in $\S$5.

%%%%%% Section2 %%%%%

\section{X-ray Observations}

\citet{2008elsner} described the Chandra X-ray observations of M71. We here note some relevant 
information for our analysis and extend the spectral fitting in this paper to test alternative models 
besides the power-laws considered by \citet{2008elsner}.
Only seven of the 29 detected X-ray sources have sufficient counts (six sources with at least 50 source counts 
and one known MSP with 37.5 source counts\footnote{The number of source counts is a result of modelling the
background and the PSF, and represents a background subtracted expectation value for the source counts 
within the detect cell.}) to warrant a detailed spectral analysis. 
We used the CIAO tool {\it dmextract} to extract spectra of the brighter sources 
and the source-free background regions near to those sources. Response files were constructed by using the 
CIAO tool {\it mkacisrmf} and {\it mkarf}. The extracted spectra were binned with at least 5 source counts per bin. 
Background-subtracted spectral modeling was performed with XSPEC 
using data in the energy band 0.3--8.0\,keV. To characterize the spectra of these sources, we fitted each of the 7 brightest 
X-ray sources with several different models (i.e., power-law (PL), thermal bremsstrahlung (TB), and blackbody (BB))
by using \citet{1979cash} statistics. Assuming all the X-ray sources within the half-mass radius are 
associated with the globular cluster M71, we fixed the hydrogen column density at the value of 
$N_\mathrm{H}=1.39 \times 10^{21}~\mathrm{cm^{-2}}$ from optical extinction to attempt spectral fitting. 

In Table~\ref{table2}, column 1 shows the Chandra source name given in \citet{2008elsner}, and column 2 lists
the spectral model we used (we omitted the models that could not provide any physically acceptable 
description of the observed spectra). 
%In our study, none of these sources can be fitted well with a blackbody model.  
Column 3 gives the minimum number of counts used to group the spectral data for fitting, 
column 4 shows the best-fit photon index ($\Gamma$) or the temperature (keV), column 5-6
gives the C-statistic and the number of PHA\footnote{PHA: Pulse Height Analysis} bins, and the last 
column lists the unabsorbed X-ray flux in units of $10^{-14}~\mathrm{ergs~s^{-1}~cm^{-2}}$ in the 
energy bands 0.3--8.0 and 0.5--2.5\,keV.  

%----------------------------------------------------------------------------------------------------------------------------
\begin{table*}
\begin{minipage}[t]{0.98\linewidth}
\caption{Spectral fits of the X-ray sources with source counts $C_\mathrm{{0.3-8.0~keV}}~\geq$ 50}  
\label{table2}               
\centering       
\renewcommand{\footnoterule}{}    
\begin{tabular}{cccccc}     
\hline\hline  
      X-ray Source &
      Model &
      Grouping\footnote{Minimum number of counts used to group the spectral data for fitting in XSPEC.} &
      $\Gamma$/kT &
      C-statistic/bins &
      $f_\mathrm{0.3-8.0}(f_\mathrm{0.5-2.5})$\footnote{Unabsorbed X-ray flux in units of $10^{-14}~\mathrm{ergs~s^{-1}~cm^{-2}}$; 
      energy bands are 0.3--8.0 (0.5--2.5)~keV.}  \\   
\hline 
s02 & PL & 8 & 3.08$\pm$0.3 & 12.11/9 & 1.49(0.74) \\
      & TB & 8 & 0.62$\pm$0.1 & 10.99/9 & 1.26(0.82) \\
%      & BB & 5 & 0.23$^{+0.01}_{-0.02}$ & 21.08/15 & 0.89(0.75) \\
s05 & PL & 30 & 1.55$\pm$0.1 & 8.20/9 & 5.67(2.25) \\
      & TB & 30 & $9.96^{+9.2}_{-3.0}$ & 10.33/9 & 5.48(2.23)\\
      & BB & 30 & 0.69$^{+0.04}_{-0.03}$ & 59.72/9 & 2.63(0.91) \\
s08 & PL & 5 & 1.70$\pm$0.5 & 2.56/6 & 0.74(0.32) \\
      & TB & 5 & 0.53 & 20.18/6 & 0.67(0.45) \\
s15 & PL & 5 & 2.57$\pm$0.3 & 9.34/11 & 1.18(0.63) \\
      & TB & 5 & $1.31^{+0.5}_{-0.3}$ & 8.23/11 & 0.84(0.57) \\
s19 & PL & 5 & 2.61$\pm$0.4 & 1.19/9 & 1.14(0.60) \\
      & TB & 5 & $1.23^{+0.7}_{-0.3}$ & 2.77/9 & 0.80(0.55) \\
s20 & PL & 8 & 1.92$\pm$0.4 & 9.52/9 & 1.71(0.83) \\
      & TB & 8 & $1.90^{+2.1}_{-0.7}$ & 8.58/9 & 1.40(0.89) \\
      & BB & 8 & $0.35^{+0.04}_{-0.03}$ & 0.84/7 & 0.83(0.72) \\
s29 & PL & 5 & 1.29$\pm$0.4 & 4.94/8 & 1.12(0.36) \\
\hline           
\end{tabular}
\end{minipage}
\end{table*}
%____________________________________________________________________________

The X-ray spectra discussed here can help us to classify the faint X-ray sources.
The brightest X-ray sources with $L_\mathrm{X} \geq 10^{32}~\mathrm{ergs~s^{-1}}$ in the energy 
band 0.5--2.5\,keV and soft spectra ($f_\mathrm{0.5-2.0~keV}/f_\mathrm{2.0-6.0~keV} \geq 1$) are 
mostly quiescent low-mass X-ray binaries \citep[qLMXBs;][]{2008verbunt}. None of the X-ray sources 
in our sample shows this characteristic, and we conclude that M71 does not contain this kind of binary systems.
Cataclysmic variables (CVs) usually have hard spectra (power-law photon indices $\Gamma<2$) and their 
X-ray luminosities are typically between a few $10^{30}$ and a few $10^{32}~\mathrm{ergs~s^{-1}}$. 
Most faint ($L_\mathrm{X} < 3\times10^{31}~\mathrm{ergs~s^{-1}}$) sources with soft spectra belong to 
chromospherically active binaries \citep[ABs;][]{2008verbunt}. 
Looking at the brightest seven X-ray sources within the half-mass radius, we see that three 
have soft spectra ($\Gamma>2$), three have hard spectra ($\Gamma<2$), and one (s20) is borderline, 
with $\Gamma\sim2$. These spectra suggest that s05, s08, and s29 might be CVs, AGN, or MSPs, 
while the softer spectra of s02, s15, s19 and s20 may indicate that these are ABs.  
Definitive classifications require optical identification, which we turn to now.

%%%%%% Section3 %%%%%

\section{Optical Observations}

\begin{figure*}
   \centering
   \includegraphics[width=13cm]{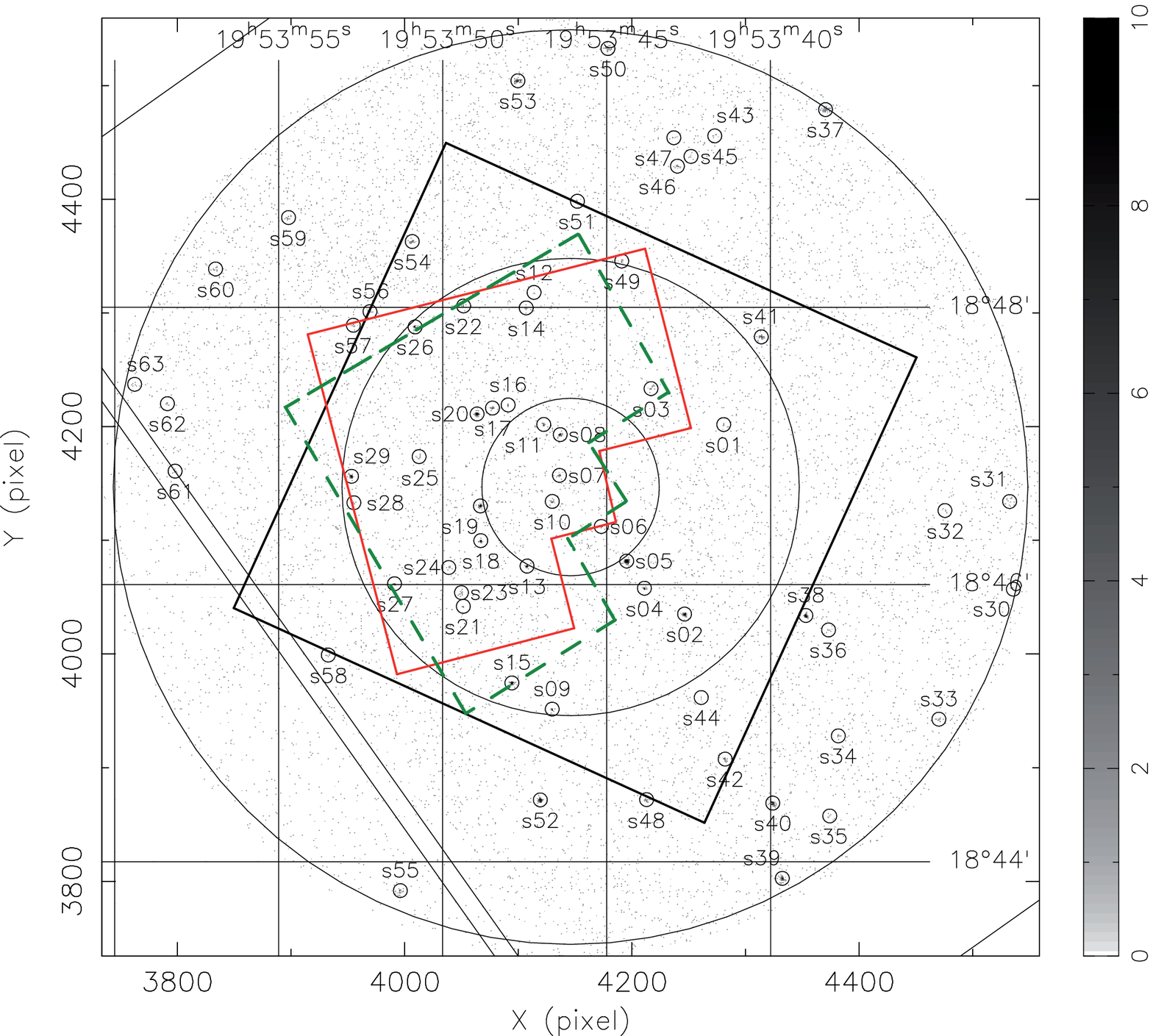}
   \caption{Chandra ACIS-S3 image of the globular cluster M71 within the energy range of 0.3 -- 8.0 keV.  
               The large circles are centered on the nominal center of the cluster and have radii $r_\mathrm{c}$ 
               (inner circle), $r_\mathrm{h}$ (middle circle), and 2$r_\mathrm{h}$ (outer circles). The small circles 
               show the positions of the 63 X-ray sources within $r_\mathrm{M71}$ $\leq$ 2$r_\mathrm{h}$ \citep{2008elsner}. 
               Straight lines mark the nominal boundaries of the ACIS-S3 and ACIS-S2 CCDs, with most of the 
               figure falling on S3 and the lower left hand portion on S2. The field of view of the HST 
               ACS (GO10775) marked by the black square covers the entire half-mass radius. The green (dashed) 
               and red (solid) polygons are the field of view of the HST WFPC2 for GO10524 and GO8118, respectively.}
    \label{fov1}%
    \end{figure*}

Two fields located inside the half-mass radius of the globular cluster M71 were observed 
with the Wide Field and Planetary Camera 2 (WFPC2) on board the Hubble Space Telescope (HST) 
in 2000 and 2006. An image of the observations is shown in Figure \ref{fov1}. For these observations, 
the PC camera was centered on the cluster center and the F336W (similar to $U$, hence 
$U_\mathrm{336}$ hereafter), F439W ($B_\mathrm{439}$), and F555W ($V_\mathrm{555}$) filters were used. 
Exposure times were 800 sec in F336W (GO10524) and 240 sec in F439W (GO8118). 
Two exposure times correspond to the F555W filter are 80 sec and 63 sec for GO10524 and GO8118, 
respectively. GO10524 also contains F255W images which did not go deep enough to identify our targets. 
To estimate whether the nondetections may be meaningful, we used \citet{2000ferraro} to estimate 
that CVs may be up to 3 magnitudes brighter (absolute magnitude) in F255W than $V$.  
Using \citet{1979seaton}, we estimate the extinction $A_\mathrm{255}=7.0$ for M71, and thus any CVs 
would be observed to be at least 3.2 magnitudes fainter in F255W than $V$.  Using the WFPC2 exposure 
time calculator (ETC), we estimate that the brightest CV candidate in our WFPC2 field, s29, could attain 
a signal-to-noise ratio (SNR) of 1.4 in the F255W data if it showed the maximum F255W/$V$ excess, 
which many of Ferraro et al.'s UV-selected objects do not.  Therefore we did not discuss the F255W 
data further in this paper. The 5-$\sigma$ limiting magnitudes of $U_\mathrm{336}$, $B_\mathrm{439}$, 
and $V_\mathrm{555}$ for point sources are 21.09, 20.87, and 21.87, respectively. 
M71 was also observed with the HST Advanced Camera for Surveys (ACS). The observations (GO10775)
consist of F606W ($V_\mathrm{606}$) and F814W ($I_\mathrm{814}$) images covering the entire half-mass 
radius of the cluster (see Fig.~\ref{fov1}). The exposure times for the F606W and F814W filters were 
304 and 324 sec with 5-$\sigma$ limiting magnitudes of $V_\mathrm{606}=25.17$ and 
$I_\mathrm{814}=23.91$ for point sources. We note that the median value of the point sources with 
the signal to noise ratio S/N $\sim$ 5 is used to define the 5-sigma limiting magnitude. 

This section outlines the data reduction, photometry, and astrometry of the HST/WFPC2 and ACS images.

%____________________________________________________________________________
%____________________________________________________________________________

\subsection{Data Reduction and Photometry}

The HST/WFPC2 data obtained from the ESO archive were processed through the 
WFPC2 Associations Science Products Pipeline\footnote{See http://archive.eso.org/archive/hst/wfpc2\_asn/}.
For each filter, single exposures were calibrated, including full bias subtraction and flat-fielding, and 
combined together in order to remove the cosmic-ray events and correct the geometrical distortions
\footnote{see http://archive.stsci.edu/hst/wfpc2/pipeline.html}.
We also downloaded the archival HST/ACS drizzled images. Those images were combined from two 
Wide Field Channel (WFC) images and calibrated with MultiDrizzle package \citep{2002koekemoer}, 
which corrected for geometric distortion and performed cosmic ray rejection.

Although M71 is a globular cluster, its stellar surface density is not as dense as that of a typical globular cluster. 
Even in the central region of M71, the average distance between stars is around $2''$, which is about 10 times larger than 
the typical FWHMs of WFPC2 and ACS cameras. Therefore, a simple aperture photometry method with the 
aperture correction is applicable to our data. We tested the flux measurement using several different psf-fitting 
photometry methods and the simple aperture photometry. We found that aperture photometry method had better 
signal-to-noise ratio and less magnitude error. Therefore, we decided to use aperture photometry to measure the fluxes
of our data.

For the data taken with WFPC2, we basically followed the instruction of aperture photometry described in 
\citet{1995holtzman}. To deal with the PSF variances within each chip and between chips, we further 
separated the images into 4 and 9 equal-size regions for the PC and WF chips, respectively, and performed 
the aperture photometry with aperture correction for each separated regions individually.
We used an aperture with the size of $0\farcs5$ in radius to measure fluxes, for all the objects with 3~$\sigma$ 
detection found using the IRAF {\it daofind} task. We note that only less than 1\% objects in each chip 
with separation of $\leq 0\farcs5$ to their neighbors, so that using an aperture with the size of $0\farcs5$ does not 
suffer from the PSF overlapping problem. The local sky values were measured using an inner sky annulus 
of 4 arcsecs with a width of 2 arcsecs, and the aperture correction value was calculated using the averages 
of the differences between the magnitudes measured using apertures with sizes of $0\farcs5$ and $4''$ in 
radius for 4 to 5 isolated stars in each separated region. The aperture correction value is 0.11$\pm$0.02 mag, 
which is consistent with the value shown in \citet{1995holtzman}. The final output magnitudes, in the 
VEGAMAG system, were corrected for the appropriate zeropoints based upon the sensitivity information 
in each header and the charge transfer efficiency effect \citep{2000dolphin}. 

For the data taken with ACS, we performed the aperture photometry based on the method described 
in \citet{2005sirianni}. The method is very similar to what we did for WFPC2. We also separated the 
ACS drizzled images into 9 equal-size regions to deal with the PSF variances. We used an aperture size 
of $0\farcs5$ in radius to measure the fluxes with sky annulus from 4 arcsecs to 6 arcsecs. The aperture 
correction value is 0.08$\pm$0.01, which agrees with the values shown in \citet{2005sirianni}. 
However, several optical counterparts, e.g., s08, s19, suffer from the PSF overlapping problem 
since the distances between them and their neighbors are $\lesssim 1\farcs0$. In order to measure accurate 
fluxes for these counterparts, we first subtracted their neighbors by using the PSF generated from isolated stars 
which are close to the counterparts, and then performed the aperture photometry on these sources. 
By doing this, we can minimize the photometric effect from the PSF wings of neighbors. 

Comparing with the photometry of M71 kindly provided by \citet{2008anderson} for the ACS images and 
reported by \citet{2002piotto} for the WFPC2 B-band and V-band images, their results are consistent with 
what we have obtained by using aperture photometry, but have the main sequence and the giant branch 
with less noise. We therefore used their photometry in our study. In addition, for those possible optical counterparts 
undetected in their photometry, we used our own results for the magnitudes which have been corrected for 
the appropriate zeropoint.
%located outside the searching region of the $2'$-radius central circle \citep{2008anderson} 
%The final output magnitudes, in the VEGAMAG system, were corrected for the appropriate zeropoints 
%based upon the sensitivity information in each header and the charge transfer efficiency effect 
%\citep{2000dolphin, 2004riess}. 

\begin{figure*}
\begin{center}
\hspace{-1.0cm}
\hbox{
\psfig{figure=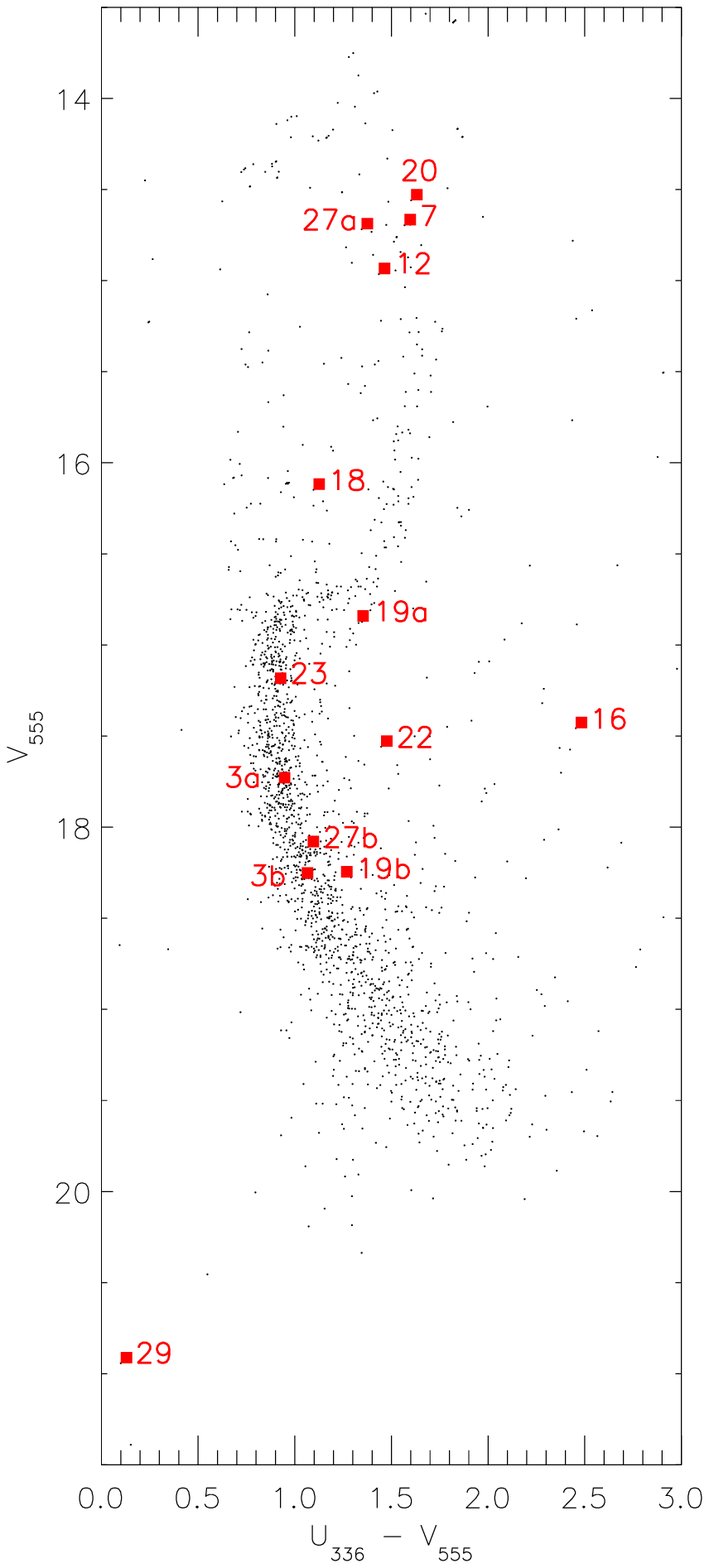,width=6.3cm}
\psfig{figure=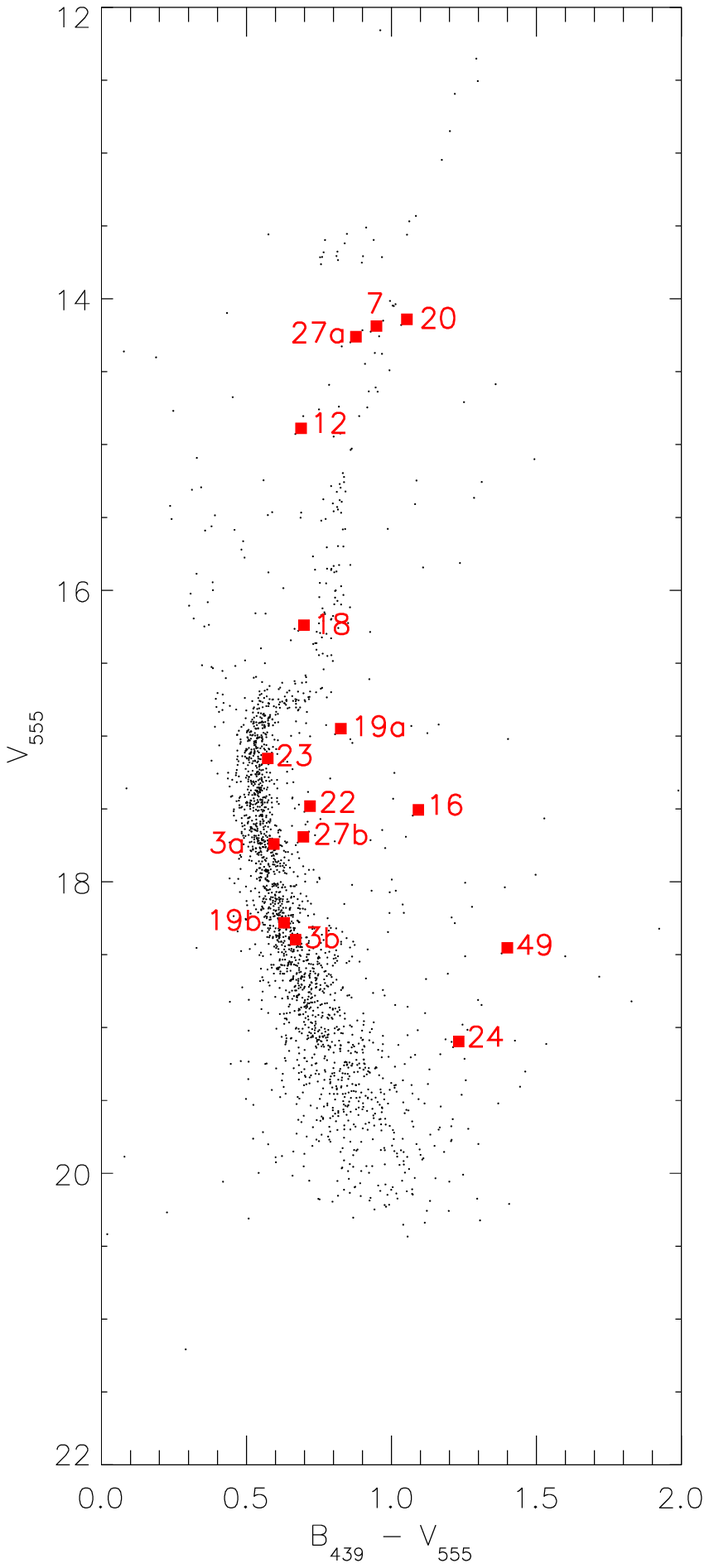,width=6.3cm}
\psfig{figure=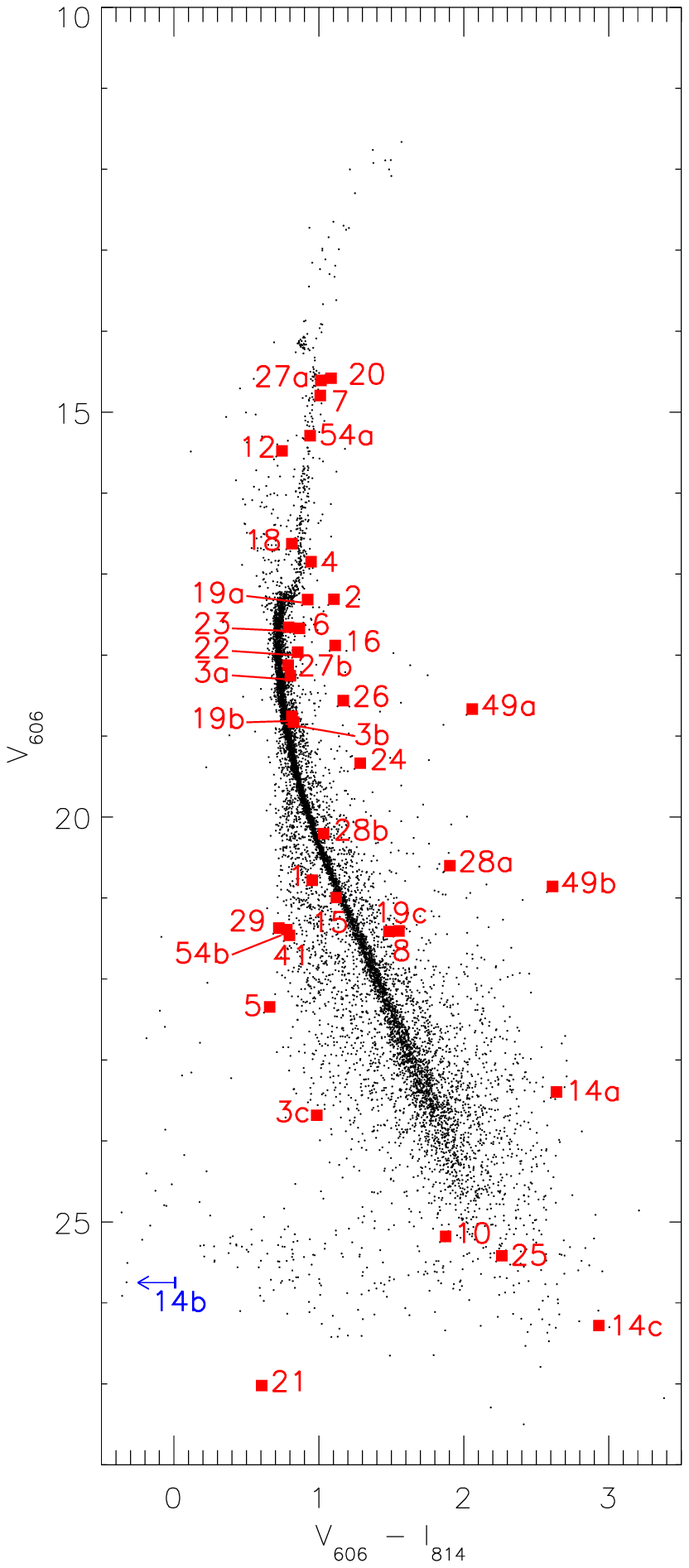,width=6.3cm}
}
\end{center}
\caption{Color magnitude diagrams (CMDs) for all the sources detected in the WFPC2 and ACS field of view.
The HST candidate counterparts matched to the X-ray sources are indicated by red squares. 
We note here s14b in the (V, V--I) CMD is plotted as a leftward-pointing arrow since its $I_\mathrm{814}$-band 
magnitude is far below the 5-sigma limiting magnitude and with a large magnitude error. }             
\label{cmd}
\end{figure*}

The most informative of these diagrams are shown in Fig.~\ref{cmd}, on which all stars located within 
the 95$\%$ confidence error circles (see Section 2 and Table 1 of Elsner et al. 2008) of the Chandra 
source positions are indicated by red squares. Numbers have been assigned to all candidate counterparts 
corresponding to the `s' designation given in \citet{2008elsner}, with `a', `b', or `c' appended if multiple potential 
optical counterparts exist.

%------------------------------------------------------------------------------------
%------------------------------------------------------------------------------------

\subsection{Astrometry}
To search for optical counterparts to the Chandra X-ray sources we aim to place both the X-ray and 
the optical frames onto the International Celestial Reference System (ICRS). We use this approach to 
improve the absolute pointing accuracy of Chandra and HST, $0\farcs6$ and $1\farcs0$ ($1\sigma$) 
respectively \citep{2000aldcroft, 2004heyer}.

For the X-ray sources, the positions listed in \citet{2008elsner} are already on the ICRS. In this paper, we aim 
to tie the HST pointing to the ICRS by finding matches between stars appearing on HST images and stars with 
accurate positions in the Two Micron All Sky Survey (2MASS) Point Source Catalog \citep{2006skrutskie}. 
On the basis of the HST pointing information contained in each image header, we used the 
WCSTools/{\it imwcs}\footnote{See http://tdc-www.harvard.edu/software/wcstools/index.html} task  on each 
corrected image to do the cross-correlation. The resulting positions were matched to those stars from the 2MASS 
catalog. There are hundreds of 2MASS stars within each HST image. By using those 2MASS stars as reference, 
the astrometric solution yielded root-mean-square residuals of $0\farcs057$ in right ascension (RA) 
and $0\farcs065$ in declination (Dec) relative to the 2MASS astrometry for the ACS images.
The resultant solution gave the residual errors of $0\farcs068$ and $0\farcs148$ in RA and $0\farcs066$ 
and $0\farcs149$ in Dec relative to the 2MASS astrometry for the PC and WF images, respectively. 
The final uncertainties of the optical source position in RA and Dec are the root of the square sum of the uncertainty 
of the astrometry in 2MASS and HST image alignment and the general uncertainties of 2MASS point source astrometry 
of typically $\sim 0\farcs1$ relative to the ICRS \citep{2006skrutskie}.
%with respect to the Tycho 2 reference system\footnote{http://pegasus.phast.umass.edu/ipac\_wget/releases/allsky/doc/ sec4\_9.html} 

%%%%%% Section4 %%%%%
\section{Source Identification and Classification}

\begin{figure*}
   \centering
   \includegraphics[width=18cm]{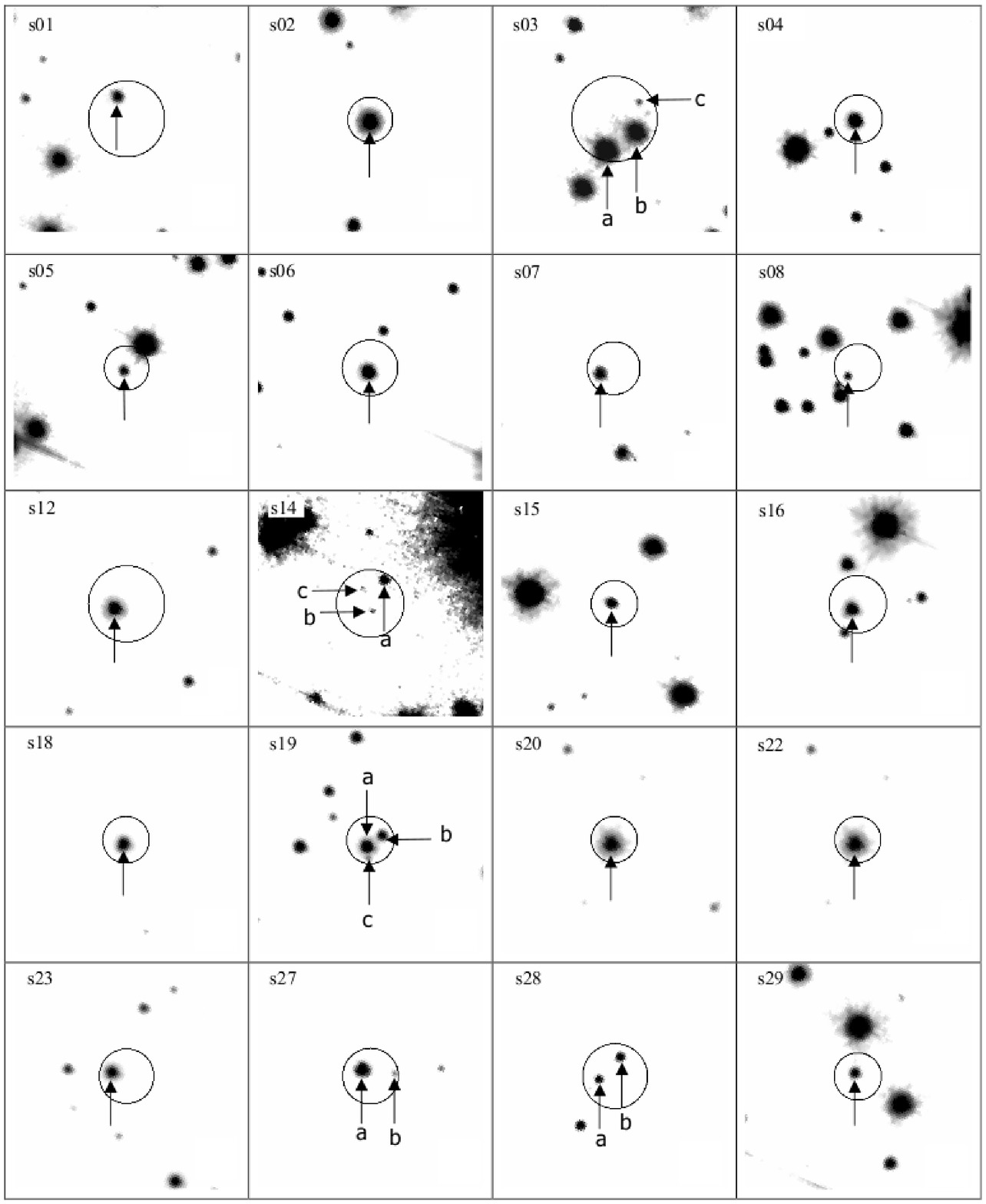}
   \vspace{-0.5cm}
   \caption{$5\arcsec\times5\arcsec$ finding charts for candidate optical counterparts within the 
    half-mass radius of M71. The finding charts are obtained from the HST/ACS $V_\mathrm{606}$ images.  
    The 95\% confidence uncertainties on the Chandra positions are overlaid on these charts, 
    while the candidate counterparts are indicated with an arrow. The finding charts are set with their dynamic range. 
    The greyscale of these images is chosen to enhance the visibility of the candidate counterparts. 
%    The possible optical counterparts, s14b, s14c, and s19c, are below the detection threshold. 
    All images have North to the top and East to the left.} 
    \label{findingchart}
    \end{figure*}

\begin{figure*}
   \centering
   \includegraphics[width=18cm]{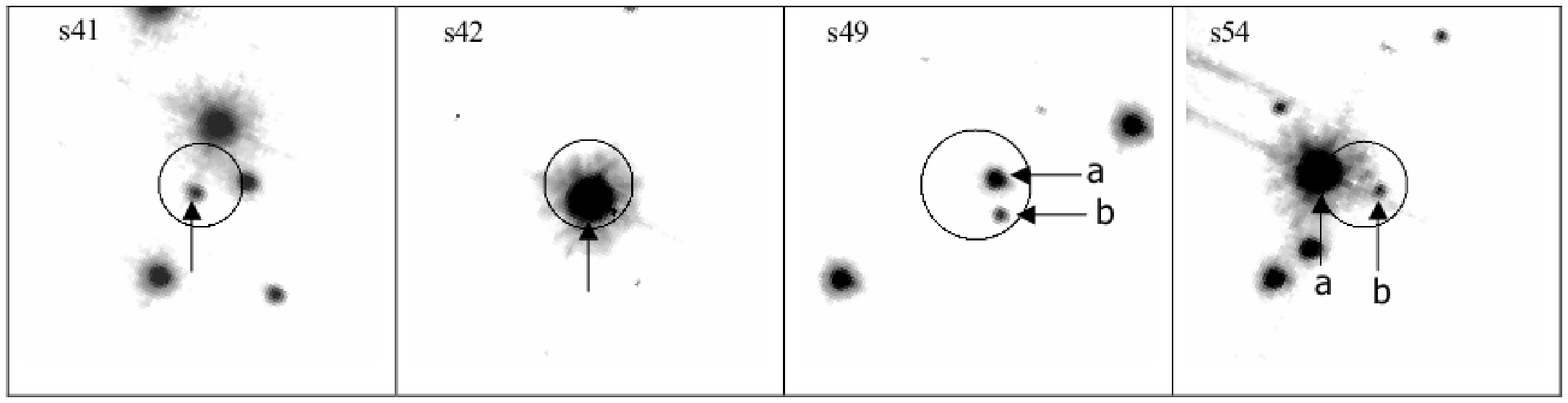}
   \vspace{-2.0cm}
   \caption{$5\arcsec\times5\arcsec$ finding charts for candidate optical counterparts outside the 
    half-mass radius of M71. The finding charts are obtained from the HST/ACS $V_\mathrm{606}$ images.  
    The 95\% confidence uncertainties on the Chandra positions are overlaid on these charts, 
    while the candidate counterparts are indicated with an arrow. The finding charts are set with their dynamic range. 
    The greyscale of these images is chosen to enhance the visibility of the candidate counterparts. 
    All images have North to the top and East to the left.} 
    \label{findingchart2}
    \end{figure*}

To obtain optical identifications for the X-ray sources we use the precise astrometry described in Section 3.
We search for optical counterparts within the 95\% Chandra error circle of the source positions
(see Table 1 of \citealt{2008elsner}), which includes the positional uncertainty of X-ray sources reported by 
the wavelet source detection algorithm, the uncertainty in the X-ray boresight correction, and the uncertainty in 
the optical astrometry. 
Within the half-mass radius of M71, there are 29 sources detected by Chandra and we suggest optical counterparts 
based on positional coincidence alone to 25 of them. In case of multiple sources inside the X-ray error circle,
we include all the candidates. The results of each candidate optical counterpart are summarized in Table~\ref{table1}, 
and finding charts are shown in Fig.~\ref{findingchart} and Fig.~\ref{findingchart2}. 

%____________________________________________________________________________
%____________________________________________________________________________

%____________________________________________________________________________
%____________________________________________________________________________

%-----------------------------------------------------------------------------------
\begin{table*}
\caption{Optical counterparts to Chandra X-ray sources within the HST/ACS field of view}   
\hspace{-0.5cm}
\begin{minipage}[t]{0.98\linewidth}          
\label{table1}      
%\centering   
\renewcommand{\footnoterule}{}        
\begin{tabular}{lrccc ccccc cc}     % 10 columns 
\hline\hline   
X-ray & \multicolumn{2}{c}{Offset From CXO position\footnote{The position offsets of sources detected in 
the $V_\mathrm{606}$ band are given relative to their Chandra positions (see Elsner et al. 2008 for details)}} & 
$U_\mathrm{336}$ & $B_\mathrm{439}$ & $V_\mathrm{555}$ & $V_\mathrm{606}$ & $I_\mathrm{814}$  & 
log($L_\mathrm{X}$)\footnote{$L_\mathrm{X}$: unabsorbed X-ray luminosity in the energy band is 0.5--2.5\,keV 
\citep[see Table 5 of][]{2008elsner}} & 
log($f_\mathrm{X}/f_\mathrm{O}$)\footnote{Ratio of X-ray to optical ($V_\mathrm{606}$) flux is computed by using 
log($f_\mathrm{X}/f_\mathrm{O}$) = log$f_\mathrm{X} + 0.4V_\mathrm{606} + 5.07$ ; $f_\mathrm{X}$ is derived in the 
0.5--2.5\,keV band.} & 
FAP\footnote{FAP: False Alarm Probability (\%). The probability of a chance coincidence in \% is calculated from 
$P_\mathrm{coinc}=N(<m)\pi r_{95}^2/A_\mathrm{search}$ \citep[see $\S$6.1 of ][]{2008elsner}. The search area, 
$A_\mathrm{search}$, is the central region of M71 with the radius of $2'$ \citep[see][]{2008anderson} and the 
considered magnitude, m, is $V_\mathrm{606}$ band magnitude. }  &
Classification\footnote{MSP: millisecond pulsar; CV: cataclysmic variable; AB: X-ray active binary; AGN: active 
galactic nucleus; F: foreground source; ?: all cases where we are not certain about the nature of the X-ray source.} \\  
\cline{2-3}  
 Source & $\Delta$RA ($''$) & $\Delta$Dec ($''$) &(mag) & (mag) & (mag) & (mag) & (mag) & $\mathrm{ergs~s^{-1}}$ & \\
\hline      
\multicolumn{12}{c}{Inside the half-mass radius} \\ 
\hline  
 s01 & 0.22 & 0.49 &  \ldots\footnote{Sources lying outside the field-of-view of the WFPC2 chips do not have 
 F336W, F439W or F555W data} & \ldots & \ldots & 20.85(0) & 19.94(2) & 29.85 & -2.02 & 26.06 & AB? \\
 s02 & 0.0 & -0.05 &  \ldots & \ldots & \ldots & 17.38(1) & 16.32(1) & 31.18 & -2.08 & 1.08 & AB \\
 s03a & 0.18 & -0.69 &  18.68(6) & 18.35(4) & 17.78(4) & 18.33(1) & 17.56(0) & 29.95 & -2.93 & 7.01 & AB? \\
 s03b & -0.50 & -0.29 &  19.32(9) & 19.09(4) & 18.44(4) & 18.90(1) & 18.11(0) & 29.95 & -2.70 & 10.36 & AB? \\
 s03c & -0.57  & 0.37 &  x\footnote{Under detection limit.} & x & x & 23.8(1) & 22.81(6) & 29.95 & -0.76 & 37.35 & CV/AGN? \\
 s04 & 0.07  & -0.02 &  \ldots & \ldots & \ldots & 16.92(1) & 16.01(1) & 30.63 & -2.82 & 0.73 & AB? \\
 s05 & 0.06 & -0.05 &  \ldots & \ldots & \ldots & 22.42(5) & 21.79(0) & 31.61 & 0.37 & 11.98 & CV/AGN? \\
 s06 & 0.04 & -0.08 &  \ldots & \ldots & \ldots & 17.74(1) & 16.91(1) & 30.15 &-2.96 & 2.67 & AB? \\
 s07 & 0.33 & -0.11 &  16.26(2) & 15.16(5) & 14.30(5) & 14.86(0) & 13.89(0) & 30.31 & -3.96 & 0.25 & AB? \\
 s08 & 0.24 & -0.17 &  x & x & x & 21.48(2) & 19.96(1) & 30.78 & -0.84 & 11.31 & MSP \\
 s10 & 0.27 &  0.05 &  x & x & x & 25.25(6) & 23.4(1) & 30.37 & 0.26 & 21.96 & CV/AGN? \\
 s12 & 0.28 & -0.10 &  16.40(2) & 15.60(3) & 14.93(3) & 15.55(0) & 14.82(0) & 29.88 & -4.11 & 0.72 & AB? \\
 s14a & -0.29 & 0.54 &  x & x & x & 23.47(6) & 20.87(2) & 30.03 & -0.80 & 33.60 & F? \\
 s14b & -0.04 & -0.18 &  x & x & x & 25.8(1) & 28(2)
% \footnote{Below the 5-$\sigma$ limiting magnitude (i.e. 23.91) in the $I_\mathrm{814}$ band.} 
   & 30.03 & 0.12 & 48.35 & CV/AGN? \\
 s14c & 0.18 & 0.36 &  x & x & x & 26.4(2)
% \footnote{Below the 5-$\sigma$ limiting magnitude (i.e. 25.17) in the $V_\mathrm{606}$ band.}  
   & 23.46(6) & 30.03 & 0.36 & 51.13 & CV/AGN? \\
 s15 & 0.10 & 0.06 &  \ldots & \ldots & \ldots & 21.06(6) & 19.98(4) & 31.0 & -0.79 & 10.37 & CV/AB?\\
 s16 & 0.15 & -0.09 &  19.9(1) & 18.62(4) & 17.55(3) & 17.95(1) & 16.88(0) & 30.15 & -2.88 & 3.44 & F? \\
% s16b & 0.29 & -0.63 &  x & 22.0(3) & 20.3(1) & 20.12(0) & 19.28(0) & 30.15 & -2.02 & 20.80 & AB? \\
 s18 & 0.08 & -0.08 & 17.25(3) & 16.96(6) & 16.28(6) & 16.69(1) & 15.92(1) & 30.77 & -2.77 & 0.58 & AB \\
 s19a & 0.08 & -0.12 & 18.20(5) & 17.79(3) & 16.99(3) & 17.39(0) & 16.50(0) & 30.95 & -2.30 & 1.16 & AB? \\
 s19b & -0.28 & 0.10 & 19.52(9) & 18.93(4) & 18.32(4) & 18.83(1) & 18.05(1) & 30.95 & -1.73 & 4.62 & AB? \\
 s19c & 0.18 & 0.36 & x & x & x & 21.48(1) & 20.04(2) & 30.95 & -0.66 & 11.33 & AB? \\
 s20 & 0.08 & -0.07 & 16.16(2) & 15.21(3) & 14.18(4) & 14.65(0) & 13.61(0) & 31.18 & -3.17 & 0.16 & AB \\ 
 s21 & 0.17 & -0.15 & x & x & x & 27.09(9) & 26.5(3) & 29.99 & 0.61 & 42.03 & CV/AGN? \\
 s22 & 0.14 & -0.15 &  19.01(7) & 18.22(5) & 17.52(3) & 18.03(0) & 17.22(1) & 30.02 & -2.98 & 5.48 & F? \\
 s23 & 0.32 & 0.09 & 18.11(5) & 17.75(3) & 17.19(3) & 17.73(1) & 16.73(1) & 30.22 & -2.91 & 2.49 & AB? \\
 s24 & 0.21 & 0.05 &  x & 20.3(1) & 19.13(4) & 19.4(1) & 18.16(1) & 30.02 & -2.43 & 9.81 & F \\
 s25 & 0.08 & 0.06 &  x & x & x & 25.49(8) & 23.3(1) & 30.03 & 0.01 & 30.34 & CV/AGN? \\
 s26 & -0.07 & -0.08 & \ldots & 19.3(2) & 18.33(7) & 18.63(7) & 17.50(3) & 30.66 & -2.10 & 4.88 & F \\
 s27a & 0.19 & 0.17 &  16.06(2) & 15.15(4) & 14.23(4) & 14.68(0) & 13.70(0) & 30.22 & -4.12 & 0.24 & AB? \\
 s27b & -0.57 & 0.06 &  19.18(8) & 18.41(3) & 17.73(3) & 18.19(1) & 17.45(1) & 30.22 & -2.72 & 4.09 & AB? \\
 s28a & 0.35 & -0.06 &  x & \ldots & \ldots & 20.67(0) & 18.81(0) & 29.99 & -1.96 & 17.96 & F? \\
 s28b & -0.15 & 0.44 &  x & \ldots & \ldots & 20.28(1) & 19.28(1) & 29.99 & -2.11 & 16.12 & AB? \\
 s29 & 0.05 & 0.06 & 21.0(2) & \ldots & 20.9(2) & 21.44(1) & 20.76(1) & 30.85 & -0.79 & 11.20 & CV \\
\hline  
\multicolumn{12}{c}{Outside the half-mass radius} \\ 
\hline
 s41 & 0.13 & -0.12 &  \ldots & \ldots & \ldots & 21.53(1) & 20.77(1) & 30.69 & -0.91 & 14.47 & CV/AGN? \\
 s49a & -0.24 & 0.09 & \ldots & 19.87(7) & 18.49(3) & 18.74(1) & 16.72(1) & 30.13 & -2.59 & 9.43 & F \\
 s49b & -0.31 & -0.38 & \ldots & x & x & 20.93(2) & 18.35(1) & 30.12 & -1.71 & 21.64 & F \\
 s54a & 0.68 & 0.20 & \dots & \dots & \ldots & 15.36(0) & 14.46(0) & 30.62 & -3.45 & 
   0.65\footnote{Since the source is outside the searching region of the $2'$-radius central 
   circle \citep{2008anderson}, the $A_\mathrm{search}$ for the probability of a chance coincidence is calculated with 
   the HST/ACS field of view, i.e. $202'' \times 202''$.} & AB? \\
 s54b & -0.23 & -0.05 & \dots & \ldots & \ldots & 21.46(1) & 20.72(1) & 30.62 & -1.01 & 19.76$^{h}$ & CV/AGN? \\  
\hline

\hline         
\end{tabular}
\end{minipage}
\end{table*}
%-----------------------------------------------------------------------------------

The first step in classifying faint ($L_\mathrm{X} \leq 10^{34.5}~\mathrm{ergs~s^{-1}}$) X-ray sources is to 
study their X-ray properties, e.g. their X-ray luminosity and spectral behavior (see $\S$2).
The second step in the identification process can be made when the sources have other information coming from different 
wavelengths. The coincidence between accurate radio timing positions of millisecond pulsars and the positions of 
X-ray sources can provide reliable identification. In the optical band, the color-magnitude diagrams (CMDs) of 
globular clusters have been studied for a long time because they reflect the fundamental properties of these 
stars and the evolutionary stage of the globular clusters. We extract further information from the locations of the 
optical stars in the CMDs of Fig.~\ref{cmd}. CVs usually lie much bluer than the main sequence stars in the (V, U--V)
and (V, B--V) CMDs while the X-ray ABs may be located on or slightly above the main sequence or 
on the giant branch. Stars on the main sequence in the V vs. V--I CMD cannot be clearly classified: they could 
either be CVs or ABs since their optical flux is dominated by the donor stars or brighter stars of the binary systems. 

\begin{figure}
\centering
\hspace{-0.8cm}
\includegraphics[width=9.5cm]{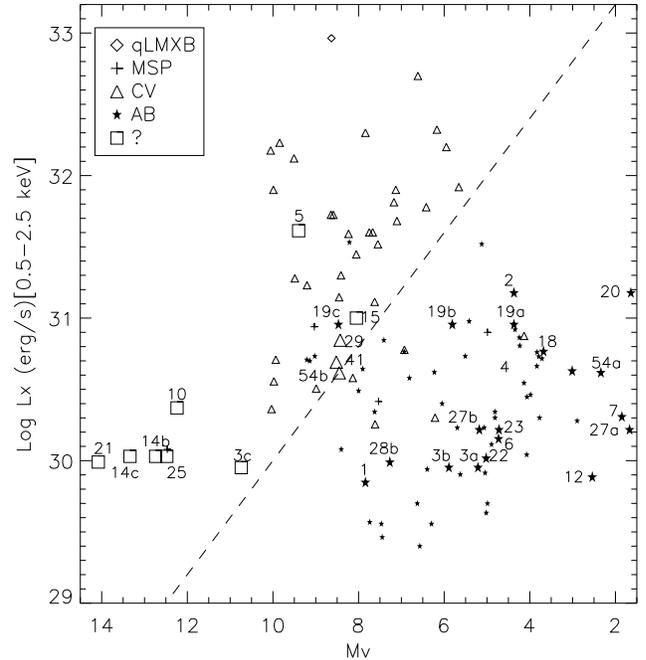}
\caption{X-ray luminosity as a function of  absolute magnitude, for low-luminosity X-ray sources 
in globular clusters. Five types of X-ray sources are shown, qLMXBs (diamonds), MSPs (crosses),
CVs (triangles), ABs (stars), and unclassified sources (squares). 
The larger and numbered symbols in this figure correspond to the optically identified X-ray sources in the
field of view of the Chandra observation of M71, where we compute absolute magnitude and X-ray 
luminosity under the assumption that the sources are associated with M71.
The smaller symbols in this figure indicate objects found in other clusters, i.e. 47\,Tuc, NGC\,6397, 
NGC\,6752, M4, NGC\,288, M55, and NGC\,6366. We note that ambiguous sources coming from other clusters 
were discarded in this figure. The dashed line of constant X-ray to optical flux ratio given by 
log $L_\mathrm{X,0.5-2.5~keV} \mathrm{(ergs~s^{-1}})$ = 34.0 -- 0.4 $M_\mathrm{V}$ 
\citep[after ][]{2004bassa} roughly separates CVs from ABs.}
\label{lxmv}
\end{figure}

The ratio of X-ray to optical flux is also useful to distinguish CVs from X-ray ABs \citep[see][]{2004bassa}. 
In Fig.~\ref{lxmv} we show the X-ray luminosity as a function of the absolute magnitude for low-luminosity 
X-ray sources from 47\,Tuc, NGC\,6397, NGC\,6752, M4, NGC\,288, M55, NGC\,6366
\citep[data from][]{2001grindlay,2003edmonds1,1998cool,2001grindlay2,2001taylor,2002pooley,2004bassa,
2006kong,2008bassa}, and M71. 
The large symbols in this figure indicate the X-ray sources with possible optical counterparts in the field of 
view of the Chandra observation of M71, while the smaller symbols show classified objects found in other clusters. 
We note that the absolute magnitudes and X-ray luminosities for the sources in the observations are computed 
under the assumption that they are cluster members. As discussed in \citet{2008elsner}, we caution 
that $\sim$ 40\% of the 29 X-ray sources within the half-mass radius are background or foreground objects.

Now we turn to those stars unrelated to the globular clusters. Foreground stars are likely to have counterparts 
not on the main sequence, have soft spectra, and have low $f_\mathrm{X}/f_\mathrm{O}$ ratios \citep{1999krautter}. 
For background active galactic nuclei (AGN), they will also have counterparts not on the main sequence, 
but with hard spectra; their $f_\mathrm{X}/f_\mathrm{O}$ ratios can be high \citep{1999krautter}. However, the 
AGN won't necessarily be detected at all; in some cases, the only object in an error circle may be a cluster 
main-sequence star that is not related to the X-ray source. 

We first consider those X-ray sources with only one suggested counterpart in the Chandra error circle.
S08 is a known millisecond pulsar, PSR J1953+1846A = M71A, with a spin period of 4.89 ms. It is in a 4.24 hr 
eclipsing binary system with a low-mass ($\geq$ 0.032 $M_{\odot}$) companion. It was discovered with Arecibo
\citep{2003ransom, 2005ransom, 2007hessels} and its radio emission is partially eclipsed in the orbital phase 
interval $0.18-0.36$ for approximately 20\% of each orbit \citep{2007hessels}. 
The X-ray counterpart was detected by \citet{2008elsner}.
Within the Chandra error circle we find a possible optical counterpart of this pulsar in the $V_\mathrm{606}$ and 
$I_\mathrm{814}$ band. The candidate optical counterpart to s08 lies on the main sequence in our V-I CMD with 
an absolute magnitude $M_\mathrm{V} \sim 8.5$, which implies that it has a mass of about 0.5\,$M_{\odot}$. 
The radio timing indicates a minimum mass of 0.03\,$M_{\odot}$ \citep{2007hessels} for the companion star. 
In order to allow for such a massive companion (i.e. $\sim$0.5\,$M_{\odot}$), conceivably the orbit could be 
seen nearly face-on (within 4 degrees). However, in that (extremely unlikely) case we do not expect regular 
radio eclipses, as are observed.
%, which seems inconsistent with this star's magnitude (stars below 0.075\,$M_{\odot}$ should have $M_\mathrm{V} < 16$, \citealt{2008richer}). 
M71A's radio properties are very similar to those of other very low-mass binary pulsars such as 
PSR J1701--3006E \citep[M62E,][]{2005freire}, and therefore we conclude that M71A's companion is not this star. 
Although this star's position agrees within $0\farcs1$ with M71A's position from radio timing (I. Stairs 2009, private comm.), 
transferred with $0\farcs1$ accuracy (1$\sigma$) onto the 2MASS frame \citep{2006skrutskie}, 
this could be coincidence due to the crowding in this field (Fig.~\ref{findingchart}); alternatively, 
M71A could be a hierarchical triple system. Future radio timing may determine this.

%This is inconsistent with the companion inferred from the radio timing, which has a mass below the hydrogen-burning 
%limit of 0.075\,$M_{\odot}$, roughly $M_\mathrm{V} = 16$ \citep{2008richer}. 
%So far, no very low-mass companion to an eclipsing MSP has been discovered in a globular cluster in the optical
%band. {\bf In addition, the position offset of the possible counterpart relative to the accurate timing solution of M71A 
%obtained from radio observations is $\sim 0\farcs02$ in RA and $\sim 0\farcs10$ in Dec (kindly provided 
%by Ingrid Stairs). Although it is comparable with 1-$\sigma$ absolute astrometric error circle inferred from the uncertainty 
%of the radio timing position and 2MASS astrometric accuracy reported by \citet{2006skrutskie} (i.e. of order 100 mas) 
%in quadrature, without any further information from the optical data, e.g. optical light curve, we could not 
%conclude that the only star within the Chandra error circle is consistent with being the optical counterpart of M71A.}
%and then suggest that it is just a chance coincidence and not related to the M71A system.

The star in the error circle of s02 is nearly located on the main-sequence turn-off point (MSTO) in the (V, V--I) CMD 
of Fig.~\ref{cmd} and slightly below the subgiant branch, which is similar to the ``red straggler'' active binaries 
seen as X-ray sources in other clusters \citep{2001albrow,2003edmonds1,2003edmonds2,2008bassa}. 
Its ratio of the X-ray to optical flux locates in the region of ABs in Fig.~\ref{lxmv}.
Since s02 has a soft spectrum and shows significant time variability in the X-ray band \citep{2008elsner} 
we suggest that s02 is a chromospherically AB and its temporal variation can be explained as flaring 
on the coronally active star. S04 and s18 are also believed to be in the same group as ABs since both 
of them are located slightly above the main-sequence turn-off point, have soft X-ray spectra, and have 
lower X-ray to optical flux ratios.

S05 is the brightest X-ray source within the half-mass radius of M71. It is worth noting that s05 is the only 
X-ray source detected with ROSAT \citep{2003panzera, 2008elsner} inside the half-mass radius. 
Its optical counterpart is bluer than the main-sequence, and it has a relatively high X-ray luminosity 
($L_\mathrm{X} \sim 4 \times 10^{31}~\mathrm{ergs~s^{-1}}$). It is unlikely to be an AB. 
Its X-ray spectrum is too hard to consider it as a quiescent low-mass X-ray binary (qLMXB). 
S05 gives a bremsstrahlung temperature consistent with $\sim$ 10 keV,
as appropriate for luminous magnetic CVs \citep{1991eracleous,2003mukai}. During the 52.4 ks observation time, 
it is not consistent with being steady at 99.9\% confidence. We suggest that a CV interpretation is plausible. 
In addition, s05 has a high value of $\mathrm{log}(f_\mathrm{X}/f_\mathrm{O}) \sim 0.37$ and the blue color, which implies 
this source could be a background AGN. However, the power-law fit of its X-ray spectrum, with photon index 
$\Gamma = 1.55 \pm 0.1$, might be considered as arising from the intra-binary shock formed due to interaction 
between the relativistic pulsar wind and material from its companion star. An irradiated main-sequence companion 
could be this blue; e.g. 47\,Tuc W \citep{2005bogdanov}. MSPs with main-sequence companions of the mass 
of $\sim 0.5 M_{\odot}$ have not yet been detected, but may well be hidden from radio detection by clouds of 
ionized gas from the companion \citep[e.g.][]{2004freire}. Therefore, we cannot rule out the interpretation that it is a 
binary MSP system though this unusual scenario must be judged unlikely.
%In addition, S05 has the highest value of $log(f_{x}/f_{o}) \sim 0.46$ among all the sources detected with Chandra. Considering this, the blue color in the (V, V-I) CMD, and the variability over the length of observation, it is possible that this source could be a background active galactic nucleus (AGN). 

The candidate cluster counterpart to s29 has ultraviolet excess with respect to the main sequence (Fig.~\ref{cmd}) 
and has a high X-ray to optical flux ratio. The source can be well fitted with a power-law model with a photon index 
of $\Gamma = 1.29 \pm 0.4$, and its X-ray luminosity is $L_\mathrm{X,0.3-8.0~keV} \sim 2.1 \times10^{31}~\mathrm{ergs~s^{-1}}$.
Its U--V color is far too blue to be an AB while the optical color is redder, almost on the main sequence.
That indicates s29 is a CV with two spectral components, a blue disk and a red companion star. 

Source s15 is a good AB candidate since there is no evidence for a blue color in the VI CMD and it has a soft X-ray 
spectrum ($f_\mathrm{0.5-2.0~keV}/f_\mathrm{2.0-6.0~keV} \geq 1$). On the other hand, without the information 
from the U--V or B--V color, a CV interpretation is still plausible. Its relatively high value of 
log\,$(f_\mathrm{X}/f_\mathrm{O}) \sim -0.79$ suggests that s15 could be a CV, although it doesn't rule out an AB.

The star in the error circle of s20 lies on the giant branch and has a soft X-ray spectrum, which gives strong 
evidence that it is a chromospherically active binary containing a giant star (i.e., a RS CVn system). 
Its temporal variation in the X-ray band \citep{2008elsner} can be explained by magnetic activity. 
Since s07 and s12 are located on the giant branch and have relatively low X-ray to optical flux ratios, 
we believe that they are likely RS CVn systems as well. 

The optical counterparts associated with those X-ray sources having lower photon statistics, s01, s06, and s23, 
are located on the main-sequence and have lower X-ray to optical flux ratios. We then consider that all of them may be 
X-ray ABs. S22 exhibits rather interesting colors. In the V--I CMD it is on the main sequence, but as we shift to 
progressively bluer colors its color gets redder and redder while in the U--V CMD it is way off the main sequence. 
Thus we suggest that it is either a foreground or background source, not associated with M71.

We turn now to the sources with more than one possible counterpart in the error circle. We find two or three possible 
optical counterparts within each of the Chandra error circles for s03, s14, s19, s27, and s28. 
S03a and s03b both fall on the main-sequence in the (V, U--V), (V, B--V), and (V, V--I) CMDs. Their colors and 
low X-ray to optical flux ratios suggest that either s03a or s03b is a chromospherically AB. 
However, the blue color and the relatively high X-ray to optical flux ratio of s03c indicates that it is a CV or 
a background AGN. S14a and s28a are located far from the main sequence, suggesting that they do not belong to M71, 
while s28b located on the main sequence could be an active binary system since it has the relatively 
low X-ray to optical flux ratio. In Fig. \ref{findingchart}, we find two additional possible optical counterparts, s14b and s14c, 
within the Chandra error circle of s14, which are fainter than the 5-$\sigma$ limiting magnitudes of $V_{\mathrm{606}}$ 
and $I_{\mathrm{814}}$. Based on the blue color and the high value of $f_\mathrm{X}/f_\mathrm{O}$, 
s14b could be either a CV or a background AGN. The optical-faint source s14c is located near the 
downward-extended part of the main sequence, suggesting that it could be a main-sequence star. 
However, it is located in the $L_\mathrm{X}$ vs. $M_\mathrm{V}$ diagram in a region where no authentic  
cluster members have been found if we compute its X-ray luminosity and absolute magnitude under the 
assumption that it belongs to M71. We then rule out the AB interpretation. Therefore, due to its high X-ray 
to optical flux ratio of this source, s14c is considered as either a good candidate for background AGN 
with a optically faint object inside the error circle not related to the X-ray source or a CV candidate with 
a secondary star that dominates the optical flux in the (V, V--I) CMD. 
There are three optical counterparts within the Chandra error circle of s19. The position of s19a is near the 
MSTO point and slightly below the subgiant branch, which is similar to the case of s02,  
while s19b is located on the main sequence. Both of their X-ray to optical flux ratios are located in the region 
that is primarily populated by ABs (Fig.~\ref{lxmv}). For the third optical counterpart, s19c, its location in the 
VI CMD and soft X-ray spectrum indicate that s19c is an active binary as well. However, without the information 
from the UV color we cannot eliminate the CV interpretation due to its relatively high X-ray to optical flux ratio, 
which is similar to the case of s15. We suggest that s19a is the most likely counterpart, 
as red stragglers are very often associated with X-ray sources \citep[e.g. ][]{2005heinke}. 
According to the positions of two possible optical counterparts to s27 in the CMDs and in the absolute 
magnitude vs. X-ray luminosity diagram, we believe that either s27a or s27b is likely to be an active binary. 

The optical counterparts of s16, s24, and s26 are located further above or to the right of the main sequence than 
the binary sequence. Hence we believe that they are foreground objects and unrelated to M71. 
The positions of two faint optical counterparts to s10 and s25 in the (V, V--I) CMD and their relatively high 
X-ray to optical flux ratios are very similar to the case of s14c so that s10 and s25 could be either CVs or 
background AGN. The highest X-ray to optical flux ratio among 39 possible counterparts and blue color 
suggests that s21 is the most likely background AGN although we cannot eliminate the interpretation of a CV. 
Furthermore, the regions of the X-ray sources, s09, s11, s13, and s17, were also 
observed with the HST/WFPC2 and ACS, but we do not find any optical counterparts inside their Chandra 
error circles. If we set the 5-$\sigma$ limiting magnitude of $V_\mathrm{606}=25.17$ as the upper limit for 
these sources, their X-ray to optical flux ratios fall on a range of $\sim$ 0.2--0.6, which are higher than the highest 
$\mathrm{log}(f_\mathrm{X}/f_\mathrm{O})$ value known for an AB in a cluster, e.g. W64 in 47\,Tuc \citep{2003edmonds1}. 
Therefore, an AB interpretation can be rejected. If we take their X-ray colors into account, s11 and s17 
are located near the bottom-right corner and close to the position of s40 in the X-ray color-color diagram shown in 
Fig. 4 of \citet{2008elsner}, which implies they have very hard spectra with over half of their counts above 2\,keV. 
This infers a high intrinsic $N_\mathrm{H}$, which strongly suggests that these are background AGN.
For the other three sources with medium X-ray colors, we then tentatively classify them all as CVs, MSPs, 
or background AGN, though AGN are probably the most likely category. 
%cannot explicitly classify them all as CVs, MSPs, or background AGN without further optical information. }
%We then tentatively classify them all as background AGN as well. 

Outside the half-mass radius of M71, we find 6 optical counterparts to X-ray sources, s41, s42, s49 and s54
in the ACS field-of-view. The candidate counterpart to s42, located on the edge of the ACS, is saturated in the 
optical band, which prevents us from obtaining a reliable magnitude of this optical source or searching for any 
other faint optical sources inside the Chandra error circle. There are 2 possible optical counterparts to s49.
Both of them are located far from the main sequence, hence we suggest that they are not associated with M71. 
Inside the error circle of s54, the brighter object s54a is on the giant branch while the fainter one, s54b, 
is located blueward of the main sequence in the CMD and has a relatively high X-ray to optical flux ratio, 
suggesting that s54b might be a CV. However, s54b lies on the spikes produced by s54a, which prevents 
us from obtaining an accurate magnitude for s54b. S41's color is bluer than the main sequence, and its X-ray 
to optical flux ratio is higher than that of an AB. We then suggest that s41 is a CV candidate. 
Furthermore, a background AGN scenario is also plausible for s41 and s54b due to their blue colors, high 
$\mathrm{log}(f_\mathrm{X}/f_\mathrm{O})$ values, and their locations outside the half-mass radius of M71, where 
they are more likely to be background sources.

%-----------------------------------------------------------------------------------
%-----------------------------------------------------------------------------------

%%%%% Section 5 %%%%%
\section{Summary and Discussion}

In summary, we find one certain CV (s29), seven possible candidate CVs (s05, s10, s14, s21, s25, s41, and s54), 
and two certain ABs (s02 and s20) and 12 good candidate ABs (s01, s03, s04, s06, s07, s12, s15, s18, 
s19, s23, s27, and s28) in the globular cluster M71. Some of our candidate CVs (and/or candidate ABs) might 
be MSPs in binary systems or AGN, which often (but not always) show blue colors.

To interpret our results, understanding how many of our objects are likely false matches will be critical.  
We calculated the expected number of false matches in several ways.  First, we shifted all our X-ray 
source positions by 9$\arcsec$ and 18$\arcsec$ (somewhat arbitrary, but chosen to be larger than 
the largest uncertainties) in four directions, and searched for matches against the $V_\mathrm{606}$ frame. 
From this exercise, we expect $11^{+4}_{-8}$ false matches among our 39 possible matches, 
indicating that $\gtrsim 70$\% of the 34 total X-ray sources in our field of view have a true match.  
By chance, then, 70\% of our false matches should occur with sources which have a true match---suggesting 
that $\sim 7.7$ sources should have two possible optical counterparts. We see seven sources that have two 
or more possible $V_\mathrm{606}$ counterparts, which is nicely consistent.

In order to know which sources are more likely to have false matches, we calculated the probability of a 
chance coincidence in \% shown in Col. 10 of Table~\ref{table1} by using Eq. (3) from \citet{2008elsner}
(see also Verbunt et al. 2008).
Within the half-mass radius, there are 6 sources with a false alarm probability smaller than 1\%, 
which indicates that those associations between X-ray sources and optical counterparts have a $>$99\% confidence level.  
Adding up the false alarm probabilities gives a total expected number of false matches of $\sim 5$, which is 
consistent with the the expectation of $11^{+4}_{-8}$ false matches above. Among the 10 X-ray sources 
with optical counterparts and $L_\mathrm{X,0.5-6.0~keV} > 4 \times 10^{30} ~\mathrm{ergs~s^{-1}}$, 
adding the false alarm probabilities indicates that roughly one of them is expected to be a false match. 
(Note that we believe s08 to be a false match, but to be a true cluster member; and that s19 has three 
potential counterparts.) 

Inside the half-mass radius of M71, we find 14 X-ray sources with 
$L_\mathrm{X,0.5-6.0~keV} > 4 \times 10^{30} ~\mathrm{ergs~s^{-1}}$, of which 10 have optical counterparts. 
Assuming that all X-ray sources outside the half-mass radius are fore- or background sources, 
we can estimate that $\sim 3.7^{+3.1}_{-1.8}$ X-ray sources among 14 are unrelated to M71. The error quote 
here is from the Poisson statistic \citep{1986gehrels}. This is consistent with our estimate above that we 
have identified true optical counterparts for $\sim 9$ X-ray sources.

\begin{figure}
\centering
\hspace{-0.3cm}
\includegraphics[angle=-90,width=9.0 cm]{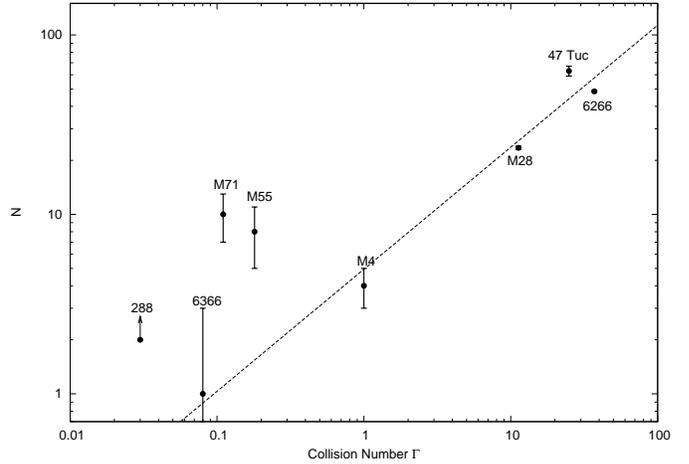}
\caption{Number of globular cluster X-ray sources (N) with $L_\mathrm{X, 0.5-6.0~keV} \geq 4 \times 10^{30} ~\mathrm{ergs~s^{-1}}$ 
inside the half-mass radius vs. the normalized collision number $\Gamma$ of the cluster. The dashed line indicates 
the best power-law fit of the data points from 47\,Tuc, NGC\,6266, M28, M4, and NGC\,6366 (cf. Table~\ref{scaling}). 
The relation between N and $\Gamma$, i.e. $\rm{N} \propto \Gamma^{0.74\pm0.03}$, is consistent with the result derived 
from \citet{2003pooley}. }
\label{gam}
\end{figure}

\citet{2003pooley} have shown that for 12 globular clusters observed by Chandra, the number of globular cluster 
X-ray sources which are above the lower limit of $4 \times 10^{30} ~\mathrm{ergs~s^{-1}}$ (0.5--6\,keV)
can be approximately linearly fitted with the predicted stellar encounter rate 
$\Gamma' \propto {\rho_o}^{1.5}{r_c}^2 \equiv \Gamma$, where $\Gamma$ is referred to as the 
collision number \citep{2003verbunt}. Here $\rho_o $ is the central density of the cluster, and $r_\mathrm{c}$ is the core radius.
In order to examine if M71 fits this relation we compare its number of X-ray sources and its collision number $\Gamma$ 
with those of some other clusters, NGC\,6266, 47\,Tuc, M28, M4, NGC\,6366, M55, and NGC\,288 (see Fig. \ref{gam}), 
using the parameters listed in Table \ref{scaling}. We note that the core-collapsed globular clusters are not 
considered in our study since their core parameters are generally uncertain, introducing strong uncertainties 
into interaction rates derived from those parameters. 
The encounter number for M71 is  $\sim$ 230 and $\sim$ 10 times smaller than those of 47\,Tuc and M4, respectively. 
\citet{2003pooley} reports 41$\pm$2 sources above the lower luminosity limit in 47\,Tuc, 
which are revised by \citet{2005heinke} to 63$\pm4$ (for a distance of 4.5 kpc); 
the uncertainty is due to the estimated number of background sources.  Thus, if the number 
of sources scales with the encounter rate, the presence of $\sim 10\pm3$ sources with 
$L_\mathrm{X,0.5-6.0~keV} > 4 \times 10^{30} ~\mathrm{ergs~s^{-1}}$ 
in M71 is a very significant overabundance, even if we take into account the errors due to Poissonian fluctuations. 
The same conclusion is reached on the basis of comparison with any other globular clusters listed in Table \ref{scaling} 
except for M55 and NGC\,288, in which the number of the X-ray sources is also in excess of the predicted value.
This indicates that most of the sources in M71, M55, and NGC\,288 are not formed via stellar encounters.

%-----------------------------------------------------------------------------------
\begin{table*}
\centering{\small
    \caption{Scaling parameters of NGC\,6266, 47\,Tuc, M28, M4, M71, NGC\,6366, M55, and NGC\,288\label{scaling}}
\begin{tabular}{lccccc cccccc}
    \hline \hline
      (1) & (2) & (3) & (4) & (5) & (6) & (7) & (8) & (9) & (10) & (11) & (12) \\
      Cluster &
      log $\rho_0$ &
      $r_\mathrm{c}$ &
      $d$ &
      $M_\mathrm{V}$ &
      $\Gamma$ &
      $M_\mathrm{h}$ &
      Source & 
      Background &
      Member & 
      IDs. &
      Reference$^{a}$ \\
      &
      ($L_\odot \mathrm{pc}^{-3}$) &
      (\arcsec) &
      (kpc) &&&&&&&&\\
\hline
    NGC\,6266  & 5.14  & 10.8 & 6.9 & $-$9.19 &  37.07 & 8.24 & 51 & 2-3 & 48-49 & 2$^{c}$ & 1,2,3 \\
    47\,Tuc       & 4.81 & 24.0 & 4.5 & $-$9.4 & 24.91 & 10.0 & 79 & $\sim$16 & 63$\pm$4 & 53-63 & 4 \\
    M\,28         &  4.75 & 14.4 & 5.6 & $-$8.33 & 11.29 & 3.73 & 26 & 2-3 & 23-24 & 2$^{c}$ & 1,5,6 \\ 
    M\,4           & 4.01 & 49.8 & 1.73 & $-$6.9 & 1.0 & 1.0 & 6 & 1-3 & 3-5 & 5 & 1,7 \\
    M\,71           & 3.05 & 37.8 & 4.0 & $-$5.6 & 0.11 & 0.30 & 14 & 1-7 & 10$\pm$3 & 4-9 & 8  \\
    NGC\,6366 & 2.42 & 109.8 & 3.6 & $-$5.77 & 0.08 & 0.33 & 5 & 2-5 & $1^{+2}_{-1}$ & $\sim$1 & 9 \\
    M\,55         & 2.15 & 169.8 & 5.3 & $-$7.6  & 0.18 & 1.82 & 16 & 5-12 & $8^{+3}_{-4}$ & 2-4 & 9 \\ 
    NGC\,288$^{b}$   & 1.80 & 85.0 & 8.4 & $-$6.7 & 0.03 & 0.83 & 11 & 4-11 & $4^{+3}_{-4}$ & 2-5 & 10 \\
\hline
\end{tabular}}
\par
\medskip
\begin{minipage}{0.98\linewidth}
      Note: Cols. 2--5 give the values for central density ($\rho_0$), core-radius
      ($r_\mathrm{c}$), distance ($d$) and absolute visual magnitude
      ($M_\mathrm{V}$) originate from \citet[ version of February 2003]{1996harris}. 
      For M4, the values of $\rho_0$ and $M_\mathrm{V}$ are computed for
      the distance and reddening of Richer et al.\,(1997). Cols. 6 and 7 are the 
      collision number which is computed from $\Gamma \equiv \rho_0^{1.5}\
      r_\mathrm{c}^2$ and the half-mass from $M_\mathrm{h} \propto
      10^{-0.4M_V}$ \citep{2006kong}. Values for $\Gamma$ and $M_\mathrm{h}$
      are normalized to the value of M4. Col. 8 shows the total number of sources detected within 
      the half-mass radius. Col. 9 is the number of expected fore/background sources. 
      Col. 10 gives the number of expected cluster members plus error.
      Col. 11 shows the number of X-ray sources (with $L_\mathrm{X,0.5-6.0~keV} > 4\times10^{30}$~ergs~s$^{-1}$) 
      which have optical or/and radio counterparts associated with the cluster, or spectrally confirmed qLMXBs.
      The last column gives the reference paper. These globular clusters are ordered 
      on the central density. \\
\noindent
a. For each cluster, the basic data for this table were extracted from: 1. \citet{2003pooley}; 2. \citet{2008cocozza}; 
3. \citet{2007trepl}; 4. \citet{2005heinke}; 5. \citet{2003becker}; 6. \citet{2007becker}; 
7. \citet{2004bassa}; 8. this work; 9. \citet{2008bassa}; 10. \citet{2006kong}. \\
b. NGC\,288 was not observed long enough to reach this luminosity limit of $L_\mathrm{X} \sim 4.0 \times 10^{30}$ 
ergs\,s$^{-1}$ in the 0.5--6.0 keV range. Its limiting luminosity is $\sim 5.7 \times 10^{30}$~ergs~s$^{-1}$. 
However, we keep this cluster's data, as a lower limit, since we have so few constraints on low-density clusters. \\
c. No information about the optical counterparts to X-ray sources.
\end{minipage}
\end{table*}

%-----------------------------------------------------------------------------------
%________________________________________________

As suggested by \citet{2002verbunt}, ABs are most likely primordial binaries, and thus to first 
order their numbers should scale with mass. Following \citet{2006kong}, we calculated the half masses 
with $10^{-0.4M_V}$, assuming the visual mass-to-light ratio is the same for all clusters listed in Table \ref{scaling}. 
M71 has the lowest half-mass, containing only $\sim$ 30\% of the mass within the half-mass radius of M\,4. 
Scaled by mass, the predicted number of ABs with $L_\mathrm{X,0.5-6~keV}> 4 \times10^{30}$~ergs~s$^{-1}$ 
in M\,71 should be similar to that of NGC\,6366 and smaller than those in any other cluster shown in Table \ref{scaling}, 
but this is in contrast to our results. 

The scaling of source number with the collision number for the sources with
$L_\mathrm{X,0.5-6~keV}>4\times10^{30}$~ergs~s$^{-1}$ suggests that CVs are mostly
made via stellar encounters \citep{2003pooley}. If we assumed that all of the CVs in M71 were formed dynamically, 
we would not expect to find more than one CV by scaling with the encounter numbers from any 
other cluster listed in Table \ref{scaling}. However, most of the globular clusters studied by \citet{2003pooley} have  
high $\Gamma$ numbers and many dynamically produced CVs. It is reasonable to suspect that 
primordial CVs may dominate in the low-density clusters. According to the computations by \citet{1997davies},  
a cluster core with a star density of $1000$~pc$^{-3}$ allows most of the CV progenitors to evolve into a CV.  
This could explain the existence of at least one (and several candidate) CVs within the half-mass radius of M71.

It is interesting to mention that in M71 $\sim$7 optical counterparts to Chandra X-ray sources are classified as 
potential RS CVn systems, in which X-ray emission is produced primarily in (sub)giant flare outbursts. 
A couple possible RS CVn X-ray sources have been identified in 47\,Tuc \citep{2005heinke} 
and $\omega$ Centauri \citep{2002cool}, but, besides M71, only the low-density clusters M55 and 
NGC 6366 \citep{2008bassa} have significant fractions of X-ray sources identified as possible RS CVns. 
Considering the even lower density case, the total X-ray luminosity of the old open cluster M\,67 is 
dominated by binaries with giants \citep{2004berg}. From the ROSAT census, the X-ray emission of most 
globular clusters per unit mass is lower than that of the old open cluster M67 \citep{2001verbunt,2002verbunt}. 
There are three possible explanations for this:
a) M67 is a very sparse cluster, which is evaporating its lowest-mass stars \citep{2005hurley}. 
Open clusters don't survive very long. As a cluster evaporates its lowest-mass stars, it tends to retain its 
heaviest systems--binaries--which are more likely to be X-ray sources.
b) M67 is a rather young cluster. Younger stars may produce more X-rays \citep{1997randich}, 
since they tend to be rotating faster than older stars.
c) A large fraction of binary systems are destroyed in globular clusters \citep[see][]{2005ivanova}, in particular those with longer orbits.  
RS CVn systems involve giants that are spun up by stellar companions. These systems must be relatively wide binaries, 
in order to avoid the giant swallowing its companion as it evolves; but such wide binaries are destroyed in globular clusters. 
Thus there are fewer RS CVn binaries in globulars. Since RS CVn binaries tend to be brighter than BY Dra binaries 
(main-sequence ABs), low-density clusters can have startlingly high X-ray luminosities per unit mass. 

To summarize the results of this paper, the number of X-ray faint sources with 
$L_\mathrm{X,0.5-6~keV} > 4\times10^{30}$~ergs~s$^{-1}$ found 
in M71 is higher than the predicted value on the basis of either the collision frequency or the half mass.
We suggest that those CVs and ABs in M71 are primordial in origin. The last interpretation above may 
explain the X-ray overabundance of low-density clusters like M\,71, where fewer primordial binaries may 
have been destroyed through binary interactions. Study of other low-density globular clusters will help us to 
better understand their evolution and dynamics.

\begin{acknowledgements}
This work made use of the Chandra and HST data archives. We acknowledge that Stairs et al. kindly provide us 
the information of M71A in advance of publication. We also thank Anderson et al. for the photometry. 
The first author thanks Albert K.H. Kong for providing some helpful suggestions and acknowledges the 
receipt of funding provided by the Max-Planck Society in the frame of the International Max-Planck Research School (IMPRS). 
COH acknowledges support from NASA Chandra grants, and funding from NSERC and the University of Alberta.   
\end{acknowledgements}

\bibliographystyle{aa}
\bibliography{bibm71}

\end{document}